# Opaque Lowermost Mantle

Sergey S. Lobanov[1,2,*], François Soubiran[3], Nicholas Holtgrewe[1], James Badro[4], Jung-Fu Lin[5], Alexander F. Goncharov[1]

[1]GFZ German Research Center for Geosciences, Section 3.6, Telegrafenberg, 14473 Potsdam, Germany

[2]Geophysical Laboratory, Carnegie Institution of Washington, Washington, DC 20015, USA

[3]École Normale Supérieure de Lyon, Université Lyon 1, Laboratoire de Géologie de Lyon, CNRS UMR5276, 69364 Lyon Cedex 07, France

[4]Université de Paris, Institut de Physique du Globe de Paris, CNRS, 75005 Paris, France

[5]Department of Geological Sciences, Jackson School of Geosciences, The University of Texas at Austin, Austin, Texas 78712, USA

*E-mail: slobanov@gfz-potsdam.de

## ABSTRACT

Earth's lowermost mantle displays complex geological structures that likely result from heterogeneous thermal and electromagnetic interaction with the core[1-4]. Geophysical models of the core-mantle boundary (CMB) region rely on the thermal and electrical conductivities of appropriate geomaterials which, however, have never been probed at representative pressure and temperature (*P*-*T*) conditions. Here we report on the opacity of single crystalline bridgmanite and ferropericlase, which is linked to both their radiative and electrical conductivity, measured in dynamically- and statically-heated diamond anvil cells as well as computed from first-principles at CMB conditions. Our results show that light absorption in the visible spectral range is enhanced upon heating in both minerals but the rate of change in opacity with temperature is a factor of six higher in ferropericlase. As a result, bridgmanite in the lowermost mantle is moderately transparent while ferropericlase is highly opaque. Our measurements suggest a very low (< 1 W/m/K) and largely temperature-independent radiative conductivity in the lowermost mantle, at odds with previous studies[5,6]. This implies that the radiative mechanism has not contributed significantly to cooling the Earth's core throughout the geologic time and points to a present-day CMB heat flow of 9-11 TW. Opaque ferropericlase is electrically conducting and mediates strong core-mantle electromagnetic coupling, explaining the intradecadal oscillations in the length of day, low secular geomagnetic variations in Central Pacific, and the preferred paths of geomagnetic pole reversals.

## MAIN TEXT

Heat flow across the CMB ($Q_{CMB}$) sustains all major geodynamic processes in the Earth's mantle and core. The observed vigor of modern plate tectonics, plume activity, and the geodynamo requires that the current $Q_{CMB}$ is 8-16 TW (Ref.[3,7]). The intensity of these geodynamic processes in the past, however, is uncertain but can be clarified if the CMB heat flow ($Q_{CMB}$) is reconstructed as a function of geologic time. An independent estimate of $Q_{CMB}$ can be obtained by using the Fourier law of heat conduction: $Q_{CMB} = A_{CMB} * k_{total} * \Delta T$ (Eq. 1), where $A_{CMB}$ is the surface area of the CMB, $\Delta T$ is the temperature gradient in the thermal boundary layer (TBL), and $k_{total}$ is the thermal conductivity of the TBL. One key unknown is the ability of the lowermost mantle to conduct heat via light radiation, which is determined by the opacity of mantle minerals at CMB *P*-*T* conditions.

The liquid outer core and the solid mantle also interact via the exchange of angular momentum, which may cause detectable variations in the Earth's rotation. Electromagnetic coupling between the core and mantle may be responsible for the reversible change in the length of day with a period of ~6 years[8] as observed by geodetic techniques. Strong coupling, however, demands that the electrical (DC) conductivity of the lower mantle minerals is sufficiently high at the CMB[9]. Furthermore, the absence of a significant lag between the rotational and magnetic signals impose a stringent limitation on the thickness of the conducting layer to be smaller than 50 kilometers[8]. Tomographic images of the lowermost mantle revealed anomalous 5-40 km thick patches directly above the core with strong seismic wave speed reductions of (~10 %), called ultra-low velocity zones (ULVZs)[4]. Because of their location just above the CMB and small thickness, these patches may be responsible for the efficient core-mantle electromagnetic coupling, yet the electrical properties of ULVZs are unknown. The DC electrical conductivity can be constrained in optical absorption experiments by extrapolating the energy-dependent optical conductivity to zero frequency. Therefore, the radiative and DC electrical conductivity can be in principle determined in a single experiment.

Insofar, the absorption coefficients of lower mantle minerals have never been measured at CMB *P*-*T* conditions. The brightness of conventional light sources is insufficient to probe hot samples with spectral radiance corresponding to several thousand degrees Kelvin and spectroscopic measurements at the conditions of combined high *P* and *T* remain a great challenge. As a consequence, information on the spectroscopic properties of mantle minerals at high *P* is largely limited to $T <$ ~1000 K. Here, we overcome the experimental limitations by employing statically- and dynamically-heated DACs coupled with laser-bright broadband pulsed optical probes and fast detectors. We report on the light absorption in single crystalline bridgmanite (Bgm), ferropericlase (Fp), and an aggregate of these minerals with realistic chemical compositions at *P*-*T* conditions representative of the lowermost mantle. We show that temperature is a major factor that governs the opacity near the base of the mantle where Bgm remains moderately transparent in the visible range while Fp is highly opaque. We reinforce our experimental findings with first-principles calculations of Fp optical properties at near CMB conditions, which constrain its absorption coefficient in the near-IR range as well as the electrical conductivity. Our results indicate extremely low radiative thermal contribution to the $Q_{CMB}$ and have profound implications to energy transport and electromagnetic coupling across the core-mantle boundary.

First, we collected high-pressure wide spectral range absorption coefficients of double-polished single crystalline Bgm6 (Bgm with 6 mol.% Fe) and Fp13 (Fp with 13 mol.% Fe) (Extended Data Fig. 1) using a conventional optical absorption setup that allows high-quality measurements at room temperature[10]. These absorption spectra reveal the distinct light absorption mechanisms that may contribute to the opacity of Bgm and Fp in the lowermost mantle. Intervalence $Fe^{2+}$-$Fe^{3+}$ charge transfer (CT) gives rise to the broad absorption band at ~17000 $cm^{-1}$ in the spectrum of Bgm6 ($Mg_{0.94}Fe^{2+}_{0.04}Fe^{3+}_{0.02}Al_{0.01}Si_{0.99}O_3$), which is close in composition to that expected for Bgm in the lower mantle[11]. Crystal field (*d*-*d*) bands were not observed in the thin (~6 µm at 117 GPa) and relatively iron-poor sample studied here, as was also the case in the previous high-pressure studies of lower mantle Bgm[6,12]. The spectrum of Fp13 showed three multiplicity-allowed low spin $Fe^{2+}$ bands. Both Bgm6 and Fp13 have a distinct UV absorption edge, typically assigned to the Fe-O CT (*e.g.* Ref.[13]).

We continued with dynamic experiments in which the samples were heated by a single 1 µs long near-infrared (1070 nm) laser pulse and probed by an ultra-bright broadband pulsed laser (Methods; Extended Data Fig. 2). Thermal radiation emitted off the dynamically-heated samples vanishes in streak camera images within ~10 µs following the arrival of the heating pulse (Extended Data Fig. 3A). Finite-element modeling of time-dependent thermal fluxes in a pulsed laser-heated DAC also indicates that ~10 µs is sufficient to restore sample's temperature back to 300 K, thanks to the high thermal conductivity of diamond[14]. Accordingly, the probe pulse train arriving with an interval of 1 µs traverses distinct thermal states and records the spectroscopic information in time domain. The timing of our dynamic experiments also allows extracting room-temperature absorption spectra prior to the arrival of the heating laser and after quenching. The obtained room-temperature spectra are in good agreement with our wide-range spectra (Fig. 1).

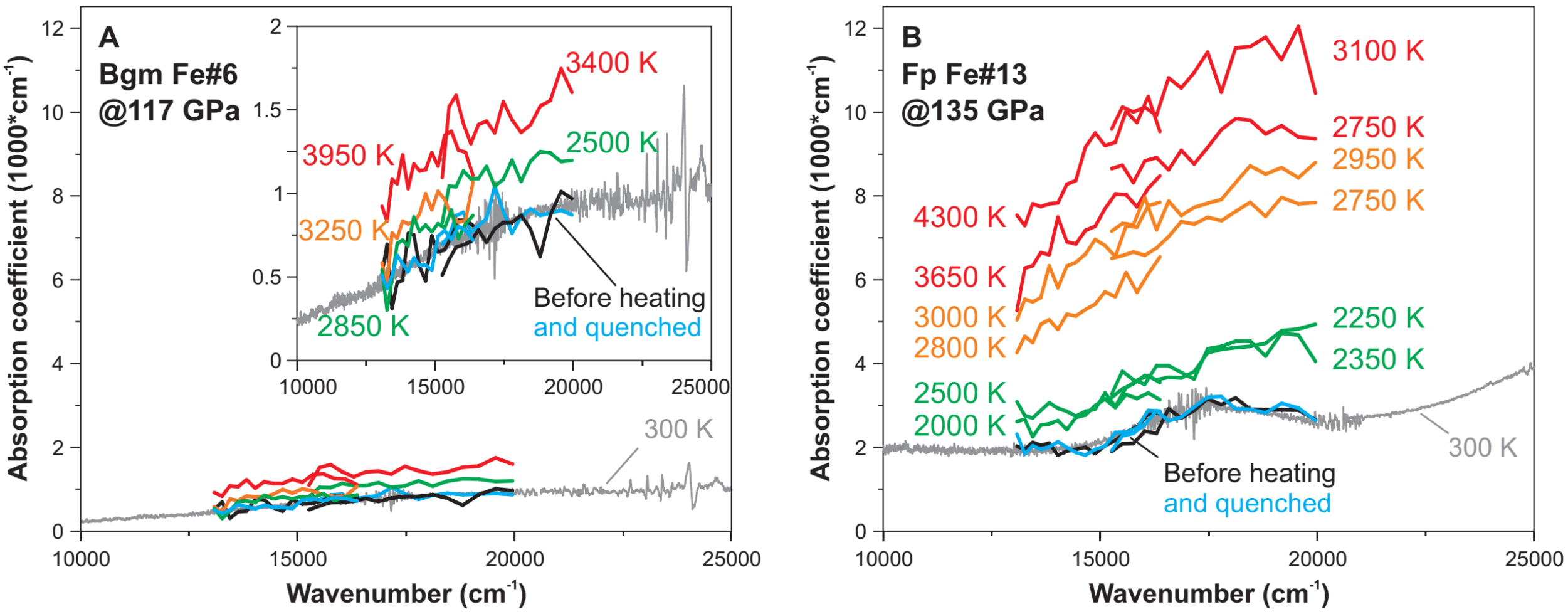

**Figure 1.** Absorption coefficients of bridgmanite at 117 GPa (**A**) and ferropericlase at 135 GPa (**B**). Black – prior to the heating pulse arrival (1-7 µs); red, orange, or green – upon cooling at high temperature (9-16 µs); and blue – after cooling (20-25 µs). Inset in (**A**) is a close-up view of Bgm6 data. Temperature uncertainty is < ± 500 K. See Methods for details. Grey spectra are absorption coefficients measured prior to heating with a conventional absorption spectroscopy setup (*e.g.* Ref.[10]). Corresponding wide-range spectra (SWIR to UV) at 300 K are shown in Extended Data Fig.1.

Upon heating of Bgm6 to ~2500 K its absorption coefficient ($\alpha$) averaged over the visible range is enhanced by approximately a factor of two, translating into a relatively small rate of increase in opacity: $\Delta\alpha/\Delta T$ of ~ 0.05 cm$^{-1}$/K (Fig. 2). At $T$ > ~3000 K, Bgm6 visible range opacity increases much more rapidly with $\Delta\alpha/\Delta T$ = 0.4 cm$^{-1}$/K, suggesting a crossover to a more efficient light absorption mechanism in Bgm across the temperature range of the TBL. Similarly, the opacity of Fp13 is enhanced at $T$ > 2000 K but with a rate that is approximately six times faster than in Bgm6 ($\Delta\alpha/\Delta T$ = 2.5 cm$^{-1}$/K). Specific absorption bands are no longer resolved in the high-temperature spectra of Bgm6 and Fp13 and the visible range opacity is evidently governed by a reversible temperature-induced red-shift of the Fe-O CT (UV absorption edge). Indeed, the initial room-temperature absorption coefficients of Bgm6 and Fp13 are restored after the samples cool down to 300 K. The reversibility in opacity over the heating cycles indicates that our pulsed laser heating time domain experiments probe intrinsic temperature-induced changes in the electronic structure as opposed to extrinsic iron redistribution due to temperature gradients in continuously laser-heated sample.

To gain quantitative information on the opacity of Bgm and Fp at $T$ < 2000 K the same DAC loadings were used for static optical absorption experiments in which the samples were continuously laser-heated for 1s and probed by the broadband pulsed laser synchronized with a gated detector (Methods). Heating of Bgm6 to ~2000 K results in a slight decrease of its $Fe^{2+}$-$Fe^{3+}$ CT band intensity while the contribution of the UV absorption edge is enhanced (Extended Data Fig. 4). This static experiment reveals the competing of individual light absorption mechanisms in Bgm6 at $T$ < 2000 K, which is the cause of the relatively small net increase of its opacity in its temperature range ($\Delta\alpha/\Delta T$ = 0.05 cm$^{-1}$/K), in excellent agreement with the rate inferred from the dynamic experiments described above (Fig. 2). Unfortunately, in static experiments on Fp13 we could not achieve satisfactory spectra reversibility at $T$ > 1000 K, which we tentatively assign to Soret-like iron diffusion due to the unavoidable temperature gradients in a laser-heated DAC. Note that the iron diffusivity in Fp is several orders of magnitude higher than in Bgm (*e.g.* Ref.[15]). Apparently, the use of a single and short laser-heating pulse in dynamic experiments allowed us to suppress this unwanted irreversible effect.

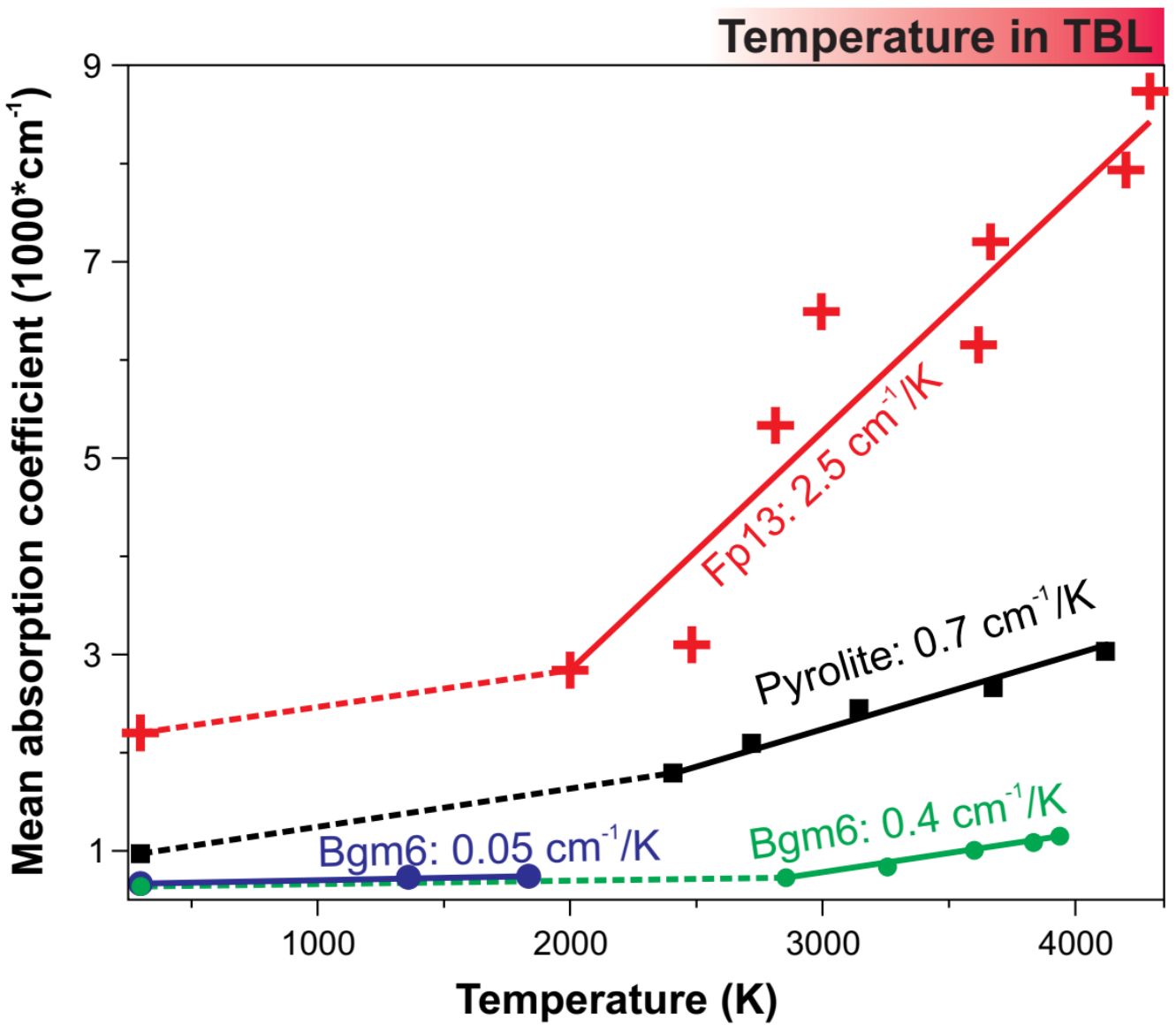

**Figure 2.** Temperature dependence of the mean absorption coefficients (13100-16400 cm$^{-1}$) observed in dynamic laser-heating experiments on bridgmanite at 117 GPa (Bgm6, green), ferropericlase at 135 GPa (Fp13, red), and pyrolite at 130 GPa (black). Dashed lines show an extrapolation from the 2500-3000 K data to 300 K. The violet

solid line shows the mean absorption coefficient of Bgm6 obtained in static laser-heating experiments. Temperature uncertainty is ~ ±500 K and ~ ±200 K in dynamic and static experiments, respectively. The red bar above the figure depicts the temperature increase expected in the thermal boundary layer (TBL).

The crossover in the slope of $\Delta\alpha/\Delta T$ in Bgm and Fp at $T > 2000$ K indicates a transition to the opacity regime dominated by the Fe-O CT, which is centered in the UV and is much more intense than *d-d* or $Fe^{2+}$-$Fe^{3+}$ transitions because electronic states of different parity (*d* and *p*) are involved in the excitation. Thus, the visible range opacity of Bgm and Fp in the lowermost mantle is governed by the Fe-O *p-d* orbital overlap. Iron in the studied Bgm6 sample is predominantly eightfold-coordinated (distorted psudododecahedral site)[11] while Fp hosts iron exclusively at the octahedral site. The *p-d* orbital overlap at the sixfold site in Fp is definitely larger than that at the twelvefold site in Bgm by virtue of a shorter Fe-O bond in Fp. As a result, the contribution of the Fe-O CT to the visible range absorbance is stronger in Fp and the corresponding $\Delta\alpha/\Delta T$ (*i.e.* temperature-induced red-shift) is a factor of six higher than in Bgm. Temperature-induced red-shifts of the Fe-O CT band have been identified in many ferromagnesian minerals at relatively low pressure and $T < 1700$ K (*e.g.* Refs.[13,16,17]), but the effect this mechanism bears on the lower mantle opacity and by extension its transport properties has never been quantified.

To understand the combined effect of Bgm and Fp on the opacity of the lower mantle in a realistic representative composition, we performed dynamic-heating optical experiments on pyrolite at 130 GPa and up to ~4000 K (Extended Data Fig. 5). We find that at $T > 2500$ K the absorption coefficient of pyrolite increases with 0.7 $cm^{-1}/K$, in excellent agreement with the expectation ($\Delta\alpha/\Delta T = 0.8$ $cm^{-1}/K$) for a 4:1 mixture of Bgm with Fp, approximating their volume fractions in a pyrolite model (Fig. 2). The derived absolute value of the mean absorption coefficient at 300 K (~1000 $cm^{-1}$) for such a pyrolite composition is sensitive to the scattering correction applied to compensate for light scattering on grain boundaries. Here, we estimated the contribution of scattering to the measured light extinction coefficient in pyrolite based on the 300 K absorption coefficients of Bgm6 and Fp13 (Extended Data Fig. 1), which is appropriate because scattering is negligible in single crystals. In any case, the extracted values of $\Delta\alpha/\Delta T$ are robust as they do not depend on the scattering correction, assuming light scattering does not change significantly with *T*. This assumption is rather accurate as values of $\Delta\alpha/\Delta T$ expected for a mixture of Bgm and Fp based on the single crystal measurements and observed directly in pyrolite are in excellent agreement. Significant grain growth over the 1 μs heating cycle, which would affect the scattering at high *T*, can also be ruled out since the temperature-enhanced absorbance of pyrolite is fully reversible (Extended Data Fig. 5).

In addition to the visible range opacity, we need to constrain the opacity in the near-IR spectra range, where most of the radiative flux is expected at all plausible mantle temperatures. Towards this end, we computed the electronic structure of $(Mg_{0.875},Fe_{0.125})O$ (Methods) at *P-T* conditions mimicking that in our optical experiments (135 GPa, 4300 K). The computed electronic density of states (DOS) shows a non-zero density of *d*-electrons at the Fermi level due to the overlapping iron *d*-orbitals (Extended Data Fig. 6). Local projection of the states identifies the peak centered at -1 eV as the $t_{2g}$ states and the peak centered at +1 eV as the $e_g$ states of iron, both mixed with oxygen *p* states. Electronic excitations between the occupied and unoccupied *d* and *p* states give rise to the distinct absorption bands observed at ~0.5 and ~2 eV (Extended Data

Fig. 7), further supporting the primary role of the Fe-O CT mechanism in the overall opacity of Fp at CMB conditions.

We model radiative thermal conductivity ($k_{rad}$) in the TBL above the CMB using the experimentally-measured absorption coefficients of Fp and Bgm at 117-135 GPa and 2500-4300 K. The measured absorption coefficients of Fp were extrapolated to 3000 $cm^{-1}$ and 25000 $cm^{-1}$ using a model that allows for a smooth decrease in the absorption coefficient with frequency (Methods, Extended Data Fig. 8A). Using this lower bound constraint on the Fp13 absorption coefficient we can now obtain its radiative thermal conductivity (Methods): ~0.2 W/m/K at 135 GP and 2500-4300 K (Fig. 3A). By extrapolating the absorption coefficients of Bgm6 in a similar fashion (Extended Data Fig. 8B) we obtain a radiative conductivity in the range of ~1.2-1.4 W/m/K at $T \sim$ 3000-4000 K (Fig. 3B). Please note that the obtained $k_{rad}$ values are upper bounds because both Fp and Bgm are expected to show absorption bands in the IR, which we did not take into account in evaluating radiative conductivity.

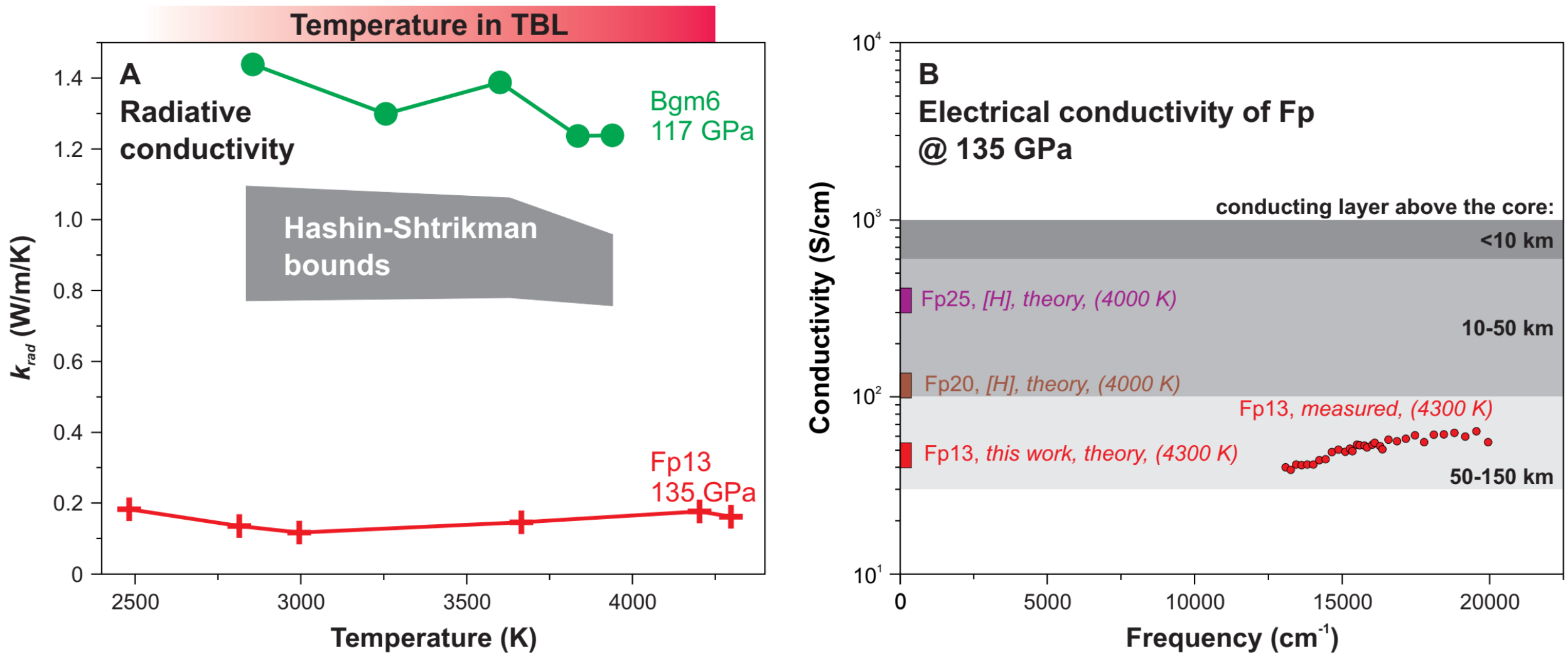

**Figure 3.** (**A**) Radiative conductivity of ferropericlase ($(Mg_{0.87},Fe_{0.13})O$) and bridgmanite ($(Mg_{0.94}Fe^{2+}_{0.04}Fe^{3+}_{0.02}Al_{0.01}Si_{0.99}O_3)$) at the *P-T* conditions of the lowermost mantle. The corresponding Hashin-Shtrikman bounds[18] for a mixture of 80 vol.% Bgm and 20 vol.% Fp are shown in black. The red bar above the figure depicts the temperature increase expected in the thermal boundary layer (TBL). (**B**) Optical conductivity of $(Mg_{0.87},Fe_{0.13})O$ measured at 135 GPa and 4300 K (red circles) and the corresponding DC electrical conductivity (red rectangle). Values for DC electrical conductivity of Fp with higher iron content from Holmstrom, et al. [19]. The grey shaded areas depict the ranges of Fp DC conductivity that would provide a conductance of $10^8$ S in the lowermost 10, 10-50, and 50-150 km when mixed with insulating Bgm (0.03 S/cm)[20], as required for the core-mantle electromagnetic coupling sufficient to produce the observed 6 year component in the length of day fluctuations[8,9].

Interestingly, radiative conductivity of Bgm and Fp at high *P-T* conditions is essentially temperature-invariant, unlike that of semi-transparent materials where $k_{rad} \sim \frac{T^3}{\alpha(P,T)}$ (Ref.[21]). Evidently, the transfer of radiative energy in the lowermost mantle is diminished by the temperature-induced opacity of Fp and Bgm revealed here. Assuming appropriate volume fractions of Bgm and Fp in the pyrolitic model (0.8 and 0.2) we obtained the Hashin-Shtrikman bounds[18] on the effective radiative conductivity in the lowermost mantle (Fig. 3A). The present results indicate that the radiative conductivity remains largely constant across the TBL and is smaller than ~1 W/m/K. The absorption coefficient of post-perovskite is about two times higher than that of Bgm at the total iron content of ~10 mol.% but shows a qualitatively similar temperature-dependence of its individual absorption bands[22] to that observed in Bgm in this work due to their crystal chemical similarity. Therefore, the inclusion of post-perovskite into the model would result in lower radiative conductivity values.

Our DFT computations also indicate that the electronic contribution to the total thermal conductivity is non-negligible and is ~ 1 W/m/K (Extended Data Fig. 9), which is consistent with the estimate of Holmstrom, et al. [19] for Fp with 19 mol.% Fe. However, the relatively small volume fraction of Fp (20 vol.%) in the lower mantle suggests that the electronic contribution of Fp to the total thermal conductivity of the lowermost mantle is insignificant (~0.2 W/m/K). Accordingly, our estimate of the total thermal conductivity of a pyrolitic mantle ($k_{total}$ = 9-11 W/m/K) only accounts for the radiative ($k_{rad}$ = 1 W/m/K, this work) and lattice contributions (8-10 W/m/K at CMB, previous studies[23-25]). In a homogeneous TBL the heat flow across the CMB is given by the Fourier law of heat conduction (Eq. 1). Accepting an average temperature gradient in TBL of ~0.007 K/m[26] and our estimate of the total thermal conductivity at the base of the mantle we obtain a $Q_{CMB}$ of 9-11 TW, which is in the range of estimates based on core energetics and mantle dynamics (8-16 TW)[3]. The apparent invariance of $k_{rad}$ to $T$ found here implies that heat transport by light radiation has remained relatively inefficient throughout geologic time and could not have promoted a higher $Q_{CMB}$ in the hotter ancient Earth.

In addition to the heat transport across the CMB, our results offer a cross-check on the geodesy-based inference of high electrical conductance ($10^8$ S) layer 10-150 km above the core. Here we showed that Bgm is insulating under near-CMB conditions as it remains relatively transparent in the visible range even at $T$ ~4000 K; thus, the potentially high DC conductivity of the lowermost mantle cannot be due to Bgm. This is also supported by previous studies that inferred a relatively low Bgm (and post-perovskite) electrical conductivity (~0.01-0.03 S/cm) at high $P$-$T$ conditions (*e.g.* Ref.[20,27]). In contrast to Bgm, the measured absorption coefficients of Fp imply that its DC conductivity is much higher than that of Bgm at near CMB conditions. The computed electrical conductivities of $(Mg_{0.875},Fe_{0.125})O$ at 135 GPa and 4300 K span ~45-165 S/cm (Extended Data Fig.10), depending mainly on the band gap correction used in the computation. This result is not only consistent with the recent theoretical estimates[19], but it falls within the range of DC conductivities required to produce the conductance of $10^8$ S in a 50-150 km thick mixture of insulating Bgm (80 vol.%) with conducting Fp (20 vol.%) (Fig. 3B). The necessary electrical conductance may be achieved even in a thin (*e.g.* < 50 km) layer just above the core if the electrical conductivity of Fp is greater than 100 S/cm. The results of this work together with previous first-principles computations[19] are consistent with such high electrical conductivity in iron-enriched Fp (> 20 mol.% Fe), which could be a plausible explanation for the six year oscillation in the length of day[8,9]. Seismic tomography images have revealed patches of ULVZs that could be explained by the occurrence of iron-enriched Fp (*e.g.* Ref.[28]). If such, these regions implement strongest core-mantle electromagnetic coupling and may manifest themselves in geomagnetic features observable at the Earth's surface. A large ULVZ located beneath the Central Pacific may electromagnetically screen the varying field of the core[1,2], which would explain the anomalously low geomagnetic secular variations observed in this region at least over the past 10-100 Ka (*e.g.* Refs.[29,30]). Likewise, electric currents in a ULVZ triggered by rapid changes in the orientation of the magnetic dipole during geomagnetic reversals may generate a torque on the core and guide the reversing dipole along the meridians that border the ULVZ (*e.g.*

Refs.[1,2]). Therefore, the preference of reversal paths that border the Pacific Ocean may be due to the ULVZ detected beneath the Pacific.

Overall, our results underscore the link between radiative and electrical conductivity. Moderately opaque and electrically insulating Bgm has small but non-negligible radiative thermal conductivity the magnitude of which determines the radiative heat flux in the lowermost mantle. Highly opaque Fp has negligible radiative thermal conductivity but its semi-metallic electrical conductivity is sufficient to implement efficient core-mantle electromagnetic coupling. Therefore, possible variations in the mineralogical abundances of these minerals along the CMB (*e.g.* in the basaltic and pyrolitic compositions) provide the means for heterogeneous CMB thermal and electromagnetic interaction. Strongest core-mantle electromagnetic interaction is expected in regions where Fp is present at the CMB, which may be detected in the secular signal of Earth's magnetic field.

**Acknowledgements**

The authors thank T. Okuchi and N. Purevjav in synthesis of Fp and Bgm samples. SSL acknowledges the support of the Helmholtz Young Investigators Group CLEAR (VH-NG-1325). FS was supported by a Marie Skłodowska-Curie action under the project ABISSE (grant agreement no. 750901). The work at Carnegie was supported by the NSF (grant numbers DMR-1039807, EAR-1520648, EAR/IF-1128867, and EAR-1763287), the Army Research Office (grant 56122-CH-H), the Deep Carbon Observatory, and the Carnegie Institution of Washington. Numerical simulations were performed on the GENCI supercomputer Occigen through the stl2816 series of eDARI computing grants. JFL acknowledges support by the NSF Geophysics Program (EAR-1446946 and EAR-1916941). JB acknowledges support by IPGP multidisciplinary program PARI, by Region Île de France SESAME Grant no. 12015908, and by Université de Paris UnivEarthS Labex program (ANR-10-LABX-0023 and ANR-11-IDEX-0005-02).

**References**

1 Runcorn, S. K. Polar Path in Geomagnetic Reversals. *Nature* **356**, 654-656 (1992).
2 Buffett, B. A. in *Treatise on Geophysics (Second Edition)* (ed Gerald Schubert) 213-224 (Elsevier, 2015).
3 Lay, T., Hernlund, J. & Buffett, B. A. Core-mantle boundary heat flow. *Nat. Geosci.* **1**, 25-32 (2008).
4 Garnero, E. J. & McNamara, A. K. Structure and dynamics of Earth's lower mantle. *Science* **320**, 626-628 (2008).
5 Goncharov, A. F., Haugen, B. D., Struzhkin, V. V., Beck, P. & Jacobsen, S. D. Radiative conductivity in the Earth's lower mantle. *Nature* **456**, 231-234 (2008).
6 Keppler, H., Dubrovinsky, L. S., Narygina, O. & Kantor, I. Optical absorption and radiative thermal conductivity of silicate perovskite to 125 Gigapascals. *Science* **322**, 1529-1532 (2008).
7 Nimmo, F. in *Treatise on Geophysics (Second Edition)* (ed Gerald Schubert) 27-55 (Elsevier, 2015).
8 Holme, R. & de Viron, O. Characterization and implications of intradecadal variations in length of day. *Nature* **499**, 202-205 (2013).
9 Buffett, B. A. Constraints on Magnetic Energy and Mantle Conductivity from the Forced Nutations of the Earth. *J. Geophys. Res. Solid Earth* **97**, 19581-19597 (1992).

10 Goncharov, A. F., Beck, P., Struzhkin, V. V., Haugen, B. D. & Jacobsen, S. D. Thermal conductivity of lower-mantle minerals. *Phys. Earth Planet. Inter.* **174**, 24-32 (2009).
11 Mao, Z. *et al.* Equation of state and hyperfine parameters of high-spin bridgmanite in the Earth's lower mantle by synchrotron X-ray diffraction and Mossbauer spectroscopy. *Am. Mineral.* **102**, 357-368 (2017).
12 Goncharov, A. F. *et al.* Experimental study of thermal conductivity at high pressures: Implications for the deep Earth's interior. *Phys. Earth Planet. Inter.* **247**, 11-16 (2015).
13 Burns, R. G. *Mineralogical applications of crystal field theory*. 2nd edn, (Cambridge University Press, 1993).
14 Montoya, J. A. & Goncharov, A. F. Finite element calculations of the time dependent thermal fluxes in the laser-heated diamond anvil cell. *J. Appl. Phys.* **111**, 112617 (2012).
15 Ammann, M. W., Brodholt, J. P. & Dobson, D. P. Ferrous iron diffusion in ferro-periclase across the spin transition. **302**, 393-402 (2011).
16 Shankland, T. J., Nitsan, U. & Duba, A. G. Optical absorption and radiative heat transport in olivine at high temperature. *J. Geophys. Res.* **84**, 1603-1610 (1979).
17 Lobanov, S. S., Holtgrewe, N. & Goncharov, A. F. Reduced radiative conductivity of low spin $FeO_6$-octahedra in $FeCO_3$ at high pressure and temperature. *Earth Planet. Sci. Lett.* **449**, 20-25 (2016).
18 Hashin, Z. & Shtrikman, S. A variational approach to theory of effective magnetic permeability of multiphase materials. *J. Appl. Phys.* **33**, 3125-3131 (1962).
19 Holmstrom, E., Stixrude, L., Scipioni, R. & Foster, A. S. Electronic conductivity of solid and liquid (Mg, Fe)O computed from first principles. *Earth Planet. Sci. Lett.* **490**, 11-19 (2018).
20 Sinmyo, R., Pesce, G., Greenberg, E., McCammon, C. & Dubrovinsky, L. Lower mantle electrical conductivity based on measurements of Al, Fe-bearing perovskite under lower mantle conditions. *Earth Planet. Sci. Lett.* **393**, 165-172 (2014).
21 Clark, S. P. Radiative transfer in the Earth's mantle. *Eos (formerly Trans. Am. Geophys. Union)* **38**, 931-938 (1957).
22 Lobanov, S. S., Holtgrewe, N., Lin, J. F. & Goncharov, A. F. Radiative conductivity and abundance of post-perovskite in the lowermost mantle. *Earth Planet. Sci. Lett.* **479**, 43-49 (2017).
23 Ohta, K., Yagi, T., Hirose, K. & Ohishi, Y. Thermal conductivity of ferropericlase in the Earth's lower mantle. *Earth Planet. Sci. Lett.* **465**, 29-37 (2017).
24 Okuda, Y. *et al.* The effect of iron and aluminum incorporation on lattice thermal conductivity of bridgmanite at the Earth's lower mantle. *Earth Planet. Sci. Lett.* **474**, 25-31 (2017).
25 Hsieh, W. P., Deschamps, F., Okuchi, T. & Lin, J. F. Effects of iron on the lattice thermal conductivity of Earth's deep mantle and implications for mantle dynamics. *Proc. Natl. Acad. Sci. U.S.A.* **115**, 4099-4104 (2018).
26 Stacey, F. D. & Davis, P. M. *Physics of the Earth*. 4th edn, (Cambridge University Press, 2008).
27 Ohta, K. *et al.* The electrical conductivity of post-perovskite in Earth's D" layer. *Science* **320**, 89-91 (2008).
28 Wicks, J. K., Jackson, J. M., Sturhahn, W. & Zhang, D. Z. Sound velocity and density of magnesiowustites: Implications for ultralow-velocity zone topography. *Geophys. Res. Lett.* **44**, 2148-2158 (2017).
29 Constable, C., Korte, M. & Panovska, S. Persistent high paleosecular variation activity in southern hemisphere for at least 10 000 years. *Earth Planet. Sci. Lett.* **453**, 78-86 (2016).
30 Panovska, S., Constable, C. G. & Korte, M. Extending Global Continuous Geomagnetic Field Reconstructions on Timescales Beyond Human Civilization. *Geochem. Geophys. Geosys.* **19**, 4757-4772 (2018).

## Extended Data and Methods

### Diamond anvil cell and sample assembly

Rhenium gaskets were indented by compression to a pressure of ~30 GPa in diamond anvil cells equipped with beveled anvils having 100/300 and 80/300 μm culets. Subsequently, circular holes with a diameter of ~50 μm were laser-drilled in the center of the indentation to serve as sample containers. After the drilling, the gaskets were washed in isopropanol for 30 min and mounted between the diamond anvils. Prior to positioning the sample, wafers of dry KCl (5 μm thick) were centered on each of the anvil. Next, double-polished single crystals of ferropericlase ($Mg_{0.87}Fe_{0.13}O$) and bridgmanite ($Mg_{0.94}Fe^{2+}_{0.04}Fe^{3+}_{0.02}Al_{0.01}Si_{0.99}O_3$) with initial thickness of ~8-10 μm were put into the sample cavity such that a sufficient area of the sample cavity was not covered by the sample to allow for reference transmission measurements through KCl (Extended Data Fig. 2A). Finally, the cells were brought to a desired pressure as gauged either by the position of the diamond Raman edge[31] or ruby fluorescence[32]. A typical discrepancy between these reading yields an ambiguity in the pressure estimate of < 5 %. No correction for thermal pressure was applied since added thermal pressure is smaller than 5 GPa at 3000 K[33,34].

### Static optical measurements at high pressure and 300 K

Here we used a custom-built all-reflective microscope combined with an IR, VIS, and near-UV conventional (non-laser) light sources. For the visible and near-UV range we used a fiber-coupled halogen-$D_2$ lamp focused to a ~50 μm spot on the sample. The transmitted portion of the radiation was collimated by a 20 μm pinhole and sent to the spectrograph (Acton Research Corporation SpectraPro 500-i) equipped with a 300 grooves/mm grating and a CCD chilled to 235 K. Measurements in the IR range were performed on the same optical bench but with a Fourier transform spectrometer equipped with a quartz beamsplitter (Varian Resolution Pro 670-IR). Details of our IR-VIS-UV setup have been reported in our previous publications[10,12,35,36]. Overall, this setup allows for a high-quality absorption spectrum in a wide spectral range (2500-30000 cm$^{-1}$) at room temperature. Absorption coefficient was evaluated as $\alpha(\nu) = \ln(10) * \frac{1}{d} * (-log_{10}(I_{sample} - I_{bckg})/(I_{reference} - I_{bckg}))$, where $d$ is sample thickness at high pressure, $I_{sample}$ is the intensity of light transmitted through the sample, $I_{reference}$ is the intensity of light passed through the KCl pressure medium, and $I_{bckg}$ is the background reading. Light losses due to the reflections at the sample-KCl interfaces are small (< 1 %) due to the similarity of the KCl and samples' refractive index at $P$ > 100 GPa ($n$ ~2) and were not taken into account.

### Static optical measurements at high pressure at $T$ < ~ 2000 K

Overall, static optical measurements at continuous laser heating allows probing the sample by a large number of probe pulses, which improves the quality of the resulting absorption spectra as compared to spectroscopic measurements in dynamic experiments (see below).

The setup combines a quasi-continuous Yt-doped 1070 nm fiber laser, a pulsed Leukos Pegasus ultra-bright supercontinuum (broadband, ~4000-25000 cm$^{-1}$) probe operating at 1 MHz, and an intensified gated CCD detector (Andor iStar SR-303i-A). The confocal probe spot size (~5 μm) was smaller than the heating laser spot (~15 μm). The spectral collection was initiated 500 ms after the start of a 1 s laser heating cycle. The detector gates were modulated for 200 ms at a rate of ~41 kHz and synchronized with the probe pulses (4 ns pulse width). Probe brightness was maximized to achieve maximum signal through the reference KCl without saturating the detector. The precise synchronization of the probe pulses and detector gates diminishes thermal background, drastically improves the signal-to-background ratio, and allows optical absorbance measurements in the VIS range (~13000-22500 $cm^{-1}$) up to ~2000 K. High-temperature absorption coefficients were evaluated as $\alpha(\nu) = \ln(10) * \frac{1}{d} * (-log_{10}(I^T_{sample} - I^T_{bckg})/(I_{reference} - I_{bckg}))$, where $I^T_{sample}$ and $I^T_{bckg}$ are the probe and background intensity at high temperature. Temperature was measured from both sides of the sample by imaging the hot spot onto the iCCD detector array. Further details of this setup can be in Lobanov, et al. [17].

### Dynamic optical measurements at high pressure and $T > \sim 2000$ K

This setup combines the same heating and probe lasers (see above) but spectral measurements were performed by a Sydor ROSS 1000 streak on a Princeton Instruments spectrometer (f/4, 150 grooves/mm). Together these components enable single-pulse laser heating experiments coupled with *in situ* time-resolved absorption measurements at $T > \sim 2000$ K[37]. Typical streak camera sweeps were 25-30 μs long and, accordingly, recorded 25-30 probe pulses each of which can be used for spectra evaluation. Importantly, spectral features and intensity of individual supercontinuum pulses are sufficiently reproducible to allow for single pulse spectroscopy (as is shown in this work). After initiation of the streak camera image collection, a single 1 μs long pulse of the 1070 nm fiber laser arrives at the $8^{th}$ μs to heat the sample (Extended Data Fig. 2B), allowing for a sufficient number of probe pulses to traverse the sample prior to heating. Sample absorption at high temperature was recorded by the streak camera images taken at two distinct grating positions centered at 700 and 590 nm, accessing 15000-20000 and 13000-16400 $cm^{-1}$ spectral ranges, respectively. From streak camera images the absorption coefficient was evaluated as: $\alpha(\nu) = \ln(10) * \frac{1}{d} * (-log_{10}(I^{time}_{sample} - I^{time}_{bckg})/(I_{reference} - I_{bckg}))$, where $I^{time}_{sample}$ and $I^{time}_{bckg}$ are the probe intensity at a given time and the corresponding (thermal) background. Similarly to the static optical experiments, reflection losses were unimportant.

Overlapping absorption spectra were stitched together to produce a spectrum in the 13000-20000 $cm^{-1}$ range (*e.g.* Fig. 1). Immediately after the collection of streak camera images the probe laser was blocked and streak camera images were measured again at identical laser heating power. These latter images were used to infer the temperature evolution of the sample for a given laser heating power. In addition, the images of clean thermal background were used to obtain $I^{time}_{bckg}$. Temperature measurements at the 700 and 590 nm grating position generally yielded consistent results. To assign temperatures to stitched spectra we relied on radiometry measurements with the grating centered at 700 nm, as more light was available for Planck fitting. However, we could only observe sufficiently intense thermal background (> 10 counts in a single streak camera sweep) at $T > \sim 3000$ K. To characterize sample absorbencies at lower temperatures, up to 100 consecutive streak camera sweeps were accumulated at low laser heating power to improve the statistics, assuming that the coupling of the sample to the heating laser did not change substantially over the 100 heating cycles. In these cases, the sample absorbance was checked afterwards to ensure its full reversibility over the heating cycles.

We estimate the overall temperature uncertainty based on the reproducibility of the absorption coefficients at high temperatures. At $T > 2000$ K, the reproducibility of the absorption coefficients was typically within 0-20 %, which translates to the overall ambiguity in the temperature measurements of < ± 500 K. This estimate is independently confirmed by optical observations of dark spots (presumably Fe-rich and formed upon melting) and increased room-temperature absorbencies in samples quenched from temperatures exceeding their expected solidus.

### Sample thickness measurements

The thickness of all studied samples was measured *ex situ* after decompression using a Zygo NewView 5032 optical 3D profilometer, which allows imaging of the surface roughness at an extremely high precision of ~10 nm. Samples were carefully extracted out of the DAC, positioned on a clean glass slide, washed with distilled water to dissolve KCl, and then brought in for Zygo imaging (Extended Data Fig. 11). The thickness at high pressure was reconstructed using the equations of state of bridgmanite ($MgSiO_3$) and periclase (MgO) assuming a perfectly elastic sample behavior upon decompression. The use of iron-free endmembers is adequate as the differences in compressibility contribute a negligibly small systematic uncertainty of < 0.2 % to the reconstructed thickness at $P$ ~ 117-135GPa.

### Radiative conductivity evaluation and Smith-Drude fitting

Under the assumption that the grain size is substantially larger than the photon mean free path the radiative conductivity of an absorbing medium is given by[21]: $k_{rad}(T) = \frac{4n^2}{3}\int_0^\infty \frac{1}{\alpha(\nu)}\frac{\partial I(\nu,T)}{\partial T}d\nu$ (Eq.2), where $\alpha(\nu)$ is the frequency-dependent absorption coefficient of the medium, $n$ its refractive index, and $I(\nu, T)$ is the Planck function. At $T$ = 3000-4000 K, the visible light photon mean free path (1/α) in Bgm and Fp is < 10 µm (Fig. 1). Accordingly, we assume that the grain size in the proximity of the core-mantle boundary is larger than 10 µm. We note that while this has been a typical assumption made in previous studies of lower mantle $k_{rad}$ [5,6,12], independent estimates of the grain size in the bulk lower mantle point towards 100-1000 µm grain size[38]; thus, validating Eq.2.

Accurate estimates of radiative conductivity require that the frequency-dependence of the absorption coefficient is known in a wide spectral range (*e.g.* 3000-25000 $cm^{-1}$). In the case of Fp, we used the Smith-Drude model[39] to extrapolate the experimental data into the IR range. Smith-Drude model only gives the lower limit on the absorption coefficient in the IR, as it does not account for *d-d* transitions, which are expected in the IR based on our theoretical computations of the Fp absorption spectrum. The following procedure was used to obtain Smith-Drude fits. First, the measured absorption coefficients ($P$ = 135 GPa, $T$ = 2500-4300 K) were converted to optical conductivity: $\sigma = n * \alpha * c * \varepsilon_0$, where $n$ is refractive index (~ 2), $c$ is the speed of light, and $\varepsilon_0$ is the permittivity of free space, which was then fit to the Smith-Drude model (with the $c$ model parameter fixed to -1, in order to gain the lowest possible values of optical conductivity in the IR). The obtained Smith-Drude optical conductivities were then converted back into the absorption coefficients (3000-25000 $cm^{-1}$ spectral range) and used to evaluate radiative conductivity at the given *P-T* conditions. Extended Data Fig. 8A shows the results of the Smith-Drude fit to the ferropericlase experimental data at 135 GPa.

In the case of Bgm, the experimentally measured absorption spectra were also extrapolated using the Smith-Drude approach outlined above. Crystal field (*d-d*) transition are expected to be important in the IR range due to the large Fe-O distance at the dodecahedral site that results in a *d-d* absorption band centered at ~7000 $cm^{-1}$ (1 atm, 300 K)[40]. Accordingly, we again expect the Smith-Drude approach to underestimate the absorption coefficient in the IR. Extended Data Fig. 8B shows the results of the Smith-Drude fit to the experimental data on bridgmanite at 117 GPa. We also tested an alternative approach to estimate the absorption coefficient of Bgm in the IR that is based on the measured rates of changes in its absorption coefficient with temperature in the visible range. Here we assume a linear and frequency-independent increase in the abortion coefficient of 0.05 and 0.4 $cm^{-1}$ at 300-3000 K and 3000-4000 K (Fig. 2), respectively. This approach always yields higher absorption coefficients in the IR than the Smith-Drude approach. However, the difference in resulting Bgm $k_{rad}$ values is relatively small: ~20 %. For the purpose of constraining the upper limit on $k_{rad}$ (Fig. 3A) we only used the Smith-Drude models to infer radiative conductivity.

**First-principles computations**

We used *ab initio* simulations to study the electronic structure of Fp and to determine the behavior of the dynamical electrical conductivity. We performed molecular dynamics of $(Mg_{0.875},Fe_{0.125})O$ coupled to density functional theory at finite temperature in the Mermin Kohn-Sham scheme[41,42] using the VASP Package[43]. The cell was composed of 28 magnesium atoms, 4 iron and 32 oxygen. We used a cubic cell with periodic boundary conditions to limit the finite size effects. We placed the atoms in a B1 structure with the iron atoms placed in lieu of magnesium atoms. The density was set to 5.48 $g/cm^3$. The temperature was controlled by a Nosé thermostat[44,45] and set to 4300 K. We used a time-step of 0.5 fs for a total duration of 7 ps. For the DFT calculation, we used projector augmented wave pseudo-potentials[46] with hard cut-offs but a frozen core of $1s^2$ for Mg and O, and $1s^2 2s^2 2p^6 3s^2$ for iron. The energy cut-off was set to 1200 eV. We used a Fermi-Dirac distribution to populate the electronic eigenstates. We sampled the Brillouin zone with the Γ-point only as it was sufficient for the trajectory. To determine the detailed electronic structure and the transport properties, we performed additional calculations on snapshots taken every 500 timestep. We used the package Abinit[47] with 672 bands and $4^3$ Monkhorst-Pack grid of K-points[48]. We used non spin-polarized calculations as the results were indistinguishable from spin polarized results. We used the linear response theory framework to study the transport properties[49,50]. We show the results of the absorption coefficient, electrical conductivity, and the electronic contribution to the total thermal conductivity as a function of excitation energy in the Extended Data Figures 7, 9 and 10. We performed calculations with the HSE06[51] functional and also with the Hubbard U correction of 2.5 eV as in Holmstrom, et al. [19].

Overall, the computed absorption coefficient of Fp at 135 GPa and 4300 K is in qualitative agreement with the experimental measurements on Fp13 but is offset to lower energy and is of a higher magnitude, likely due to the underestimated band gap, which is a known issue of PBE-DFT[50]. In order to correct for the gap underestimation that is common with GGA-DFT, we manually shifted the eigenenergies of the bands above the Fermi level by a fixed value of +0.5 or +1.0 eV[52]. Nevertheless, this shift is not fully consistent as the Kohn-Sham orbitals are likely to be modified by this operation and this was not taken into account. By rescaling the magnitude of the absorption coefficient we obtained a decent agreement with the experimental results as can be seen in Extended Data Figure 7.

Regardless of the used corrections, the inference of strongly absorbing Fp in the IR range is robust as Holmstrom, et al. [19] found largely similar DOS of $(Mg_{0.75},Fe_{0.25})O$ at comparable *P-T* conditions using a DFT methodology that included a Hubbard correction. Also, related orbital overlaps have been found in FeO at finite temperature using DFT coupled to dynamic mean field theory (DMFT) calculations[53]. This similarity indicates that Mg does not prevent the gap closure in the iron system. It also gives us confidence on the accuracy of our *ab initio* results despite the lack of DMFT formalism in our case.

## Extended Data Figures

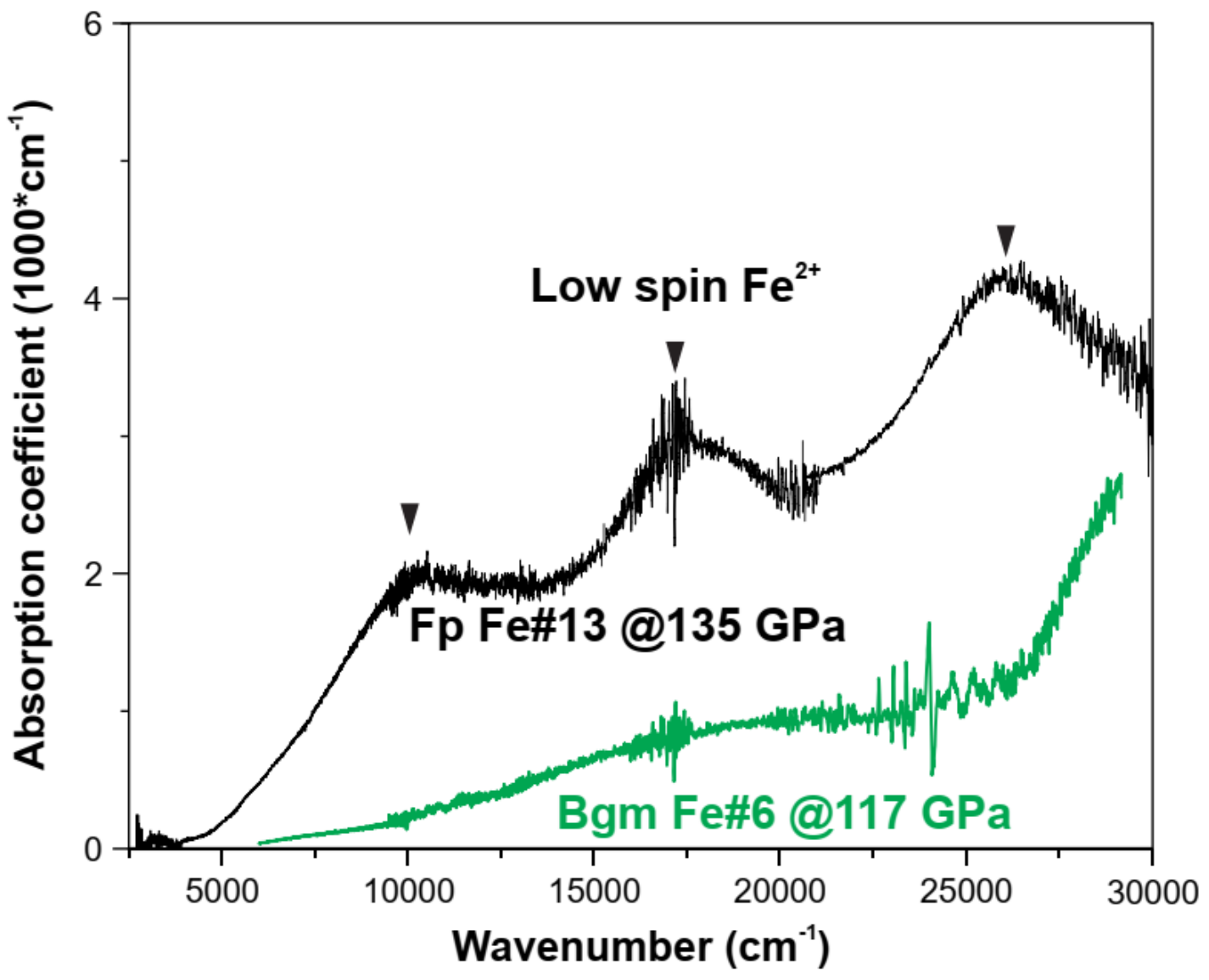

**Extended Data Figure 1.** Room-temperature absorption coefficients of Bgm6 (green) and Fp13 (black) at 117 and 135 GPa, respectively, probed by a conventional optical spectroscopy in infrared, visible, and near-ultraviolet spectral ranges[10,12].

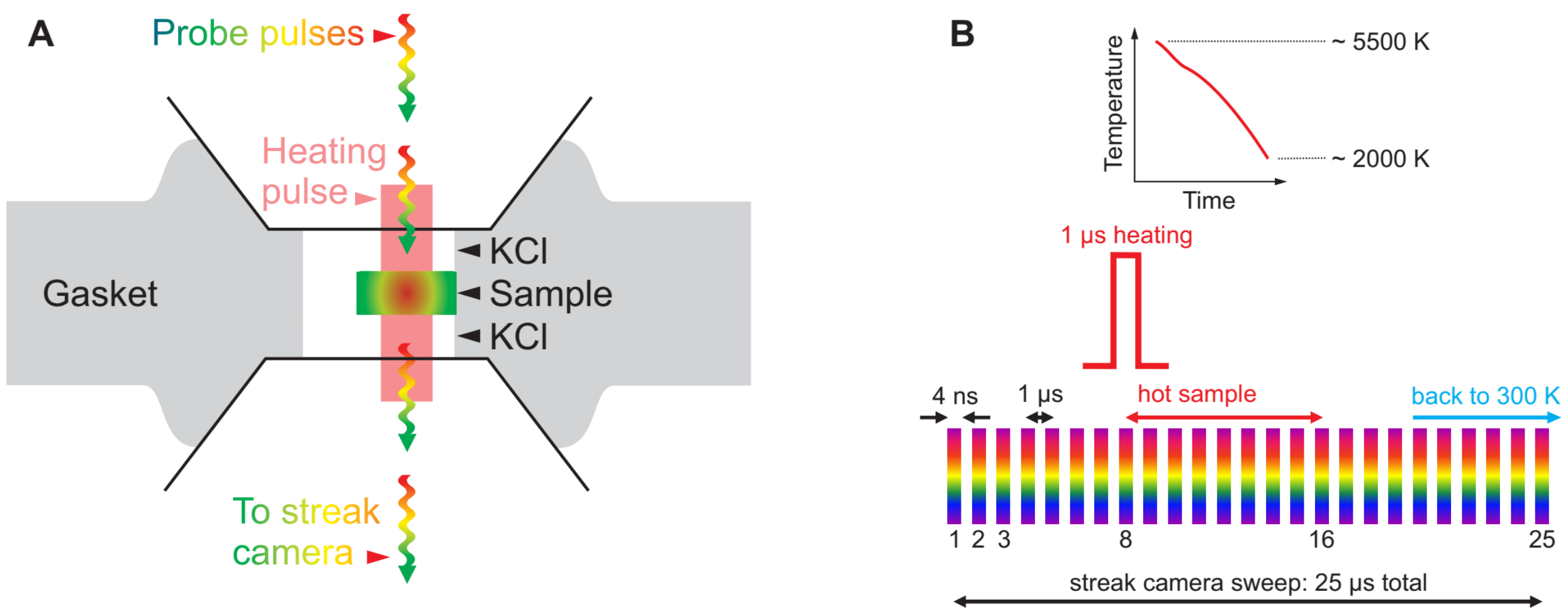

**Extended Data Figure 2.** (**A**) Diamond anvil cell assemblage used in this work. Samples were sandwiched between two KCl wafers and positioned in the cavity such that part of it can be used to measure optical reference (through KCl only). (**B**) Timing of our single laser-heating shot experiments. Probe pulses (supercontinuum laser) traverse the sample every 1 µs. The 1 µs heating laser (1070 nm, double-sided) arrives at 8 µs of the 25-30 µs long streak camera sweep.

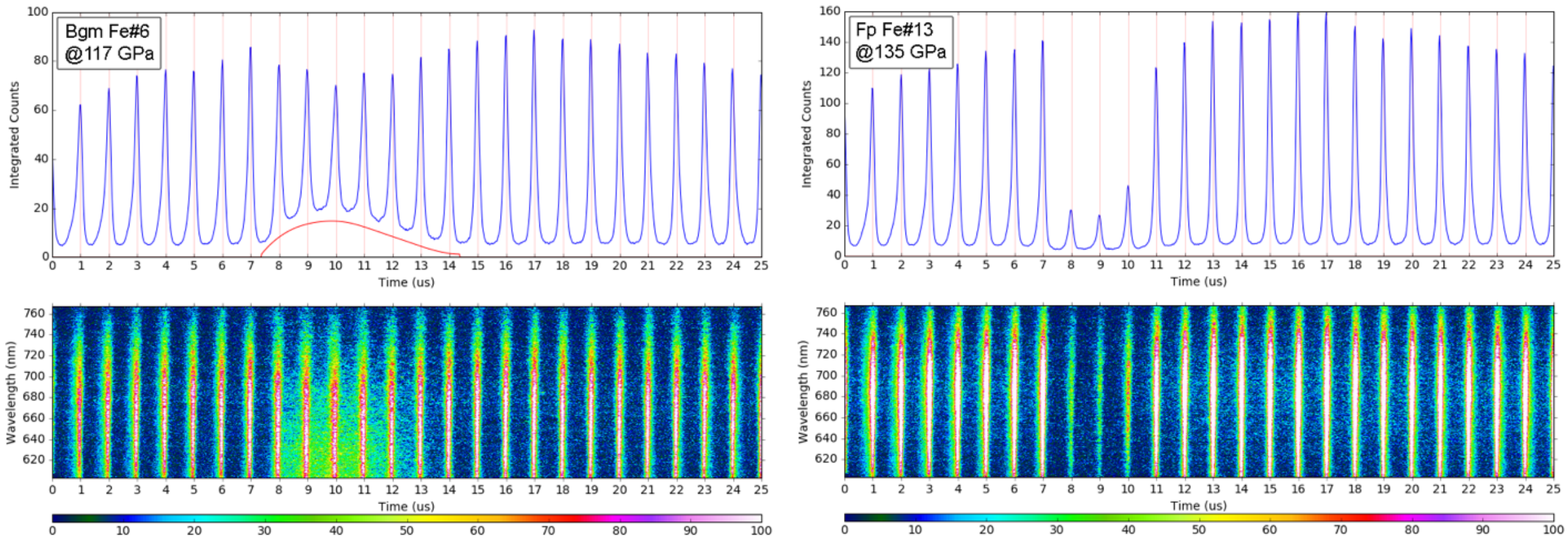

**Extended Data Figure 3.** Representative streak camera images (bottom panels) and corresponding integrated intensity (top panels) of Bgm6 at 117 GPa (left) and Fp13 at 135 GPa (right). The 1 µs laser heating pulse arrived at ~8$^{th}$ microsecond heating the samples to a maximum temperature of ~4000 K (Bgm) and ~3000 K (Fp), in these particular shots. Note the presence of apparent thermal background in the case of bridgmanite (top panel, red curve).

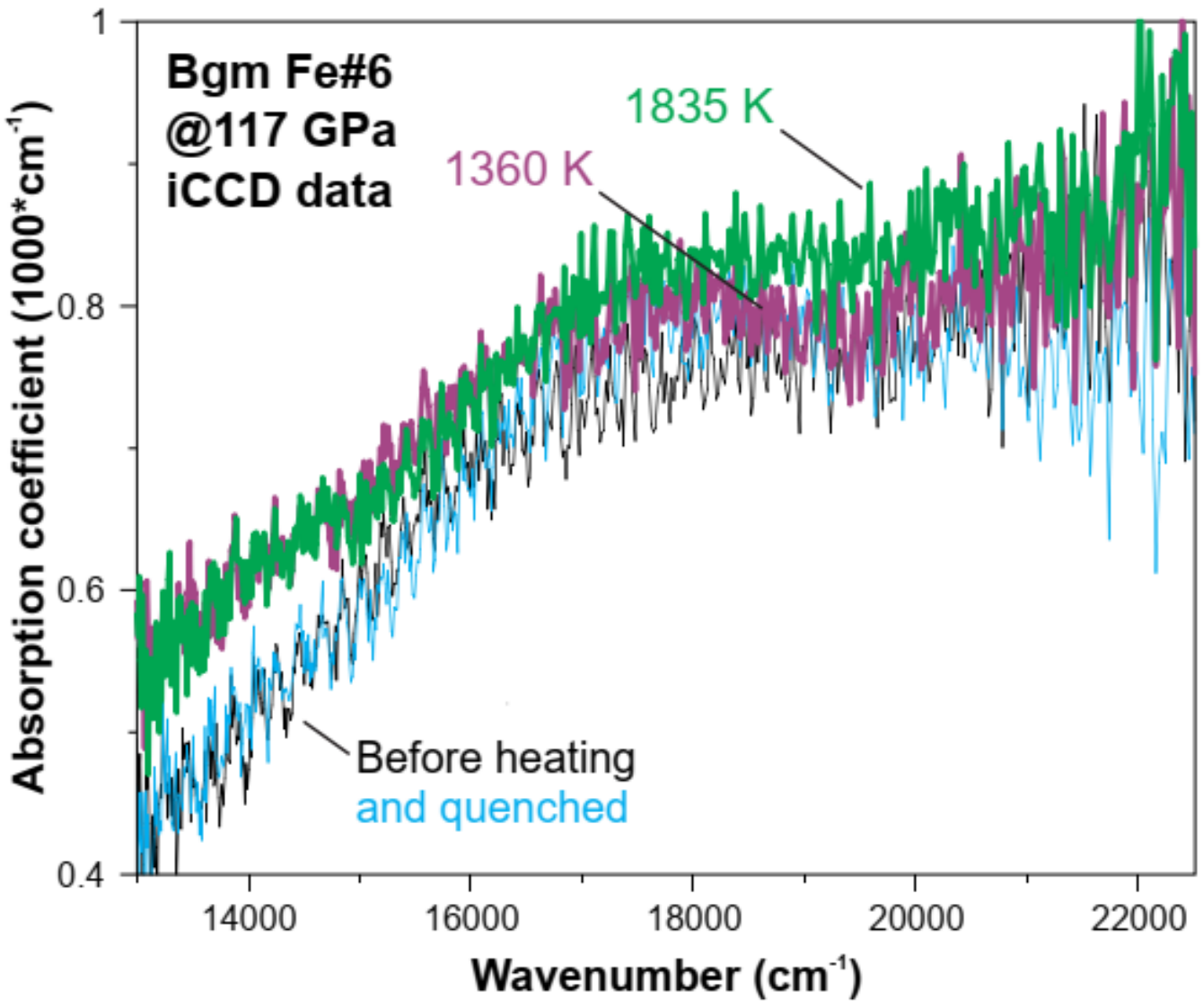

**Extended Data Figure 4.** Absorption coefficient of bridgmanite with 6 mol.% Fe at 117 GPa before and after heating (black and blue), 1360 K (purple), and 1835 K (green). The temperature dependence of the absorption coefficient is frequency-dependent. Mean absorption coefficient (averaged over the shown spectral) increases with temperature as ~0.05 cm$^{-1}$/K. Detailed description of the used experimental setup is provided in Ref.[17].

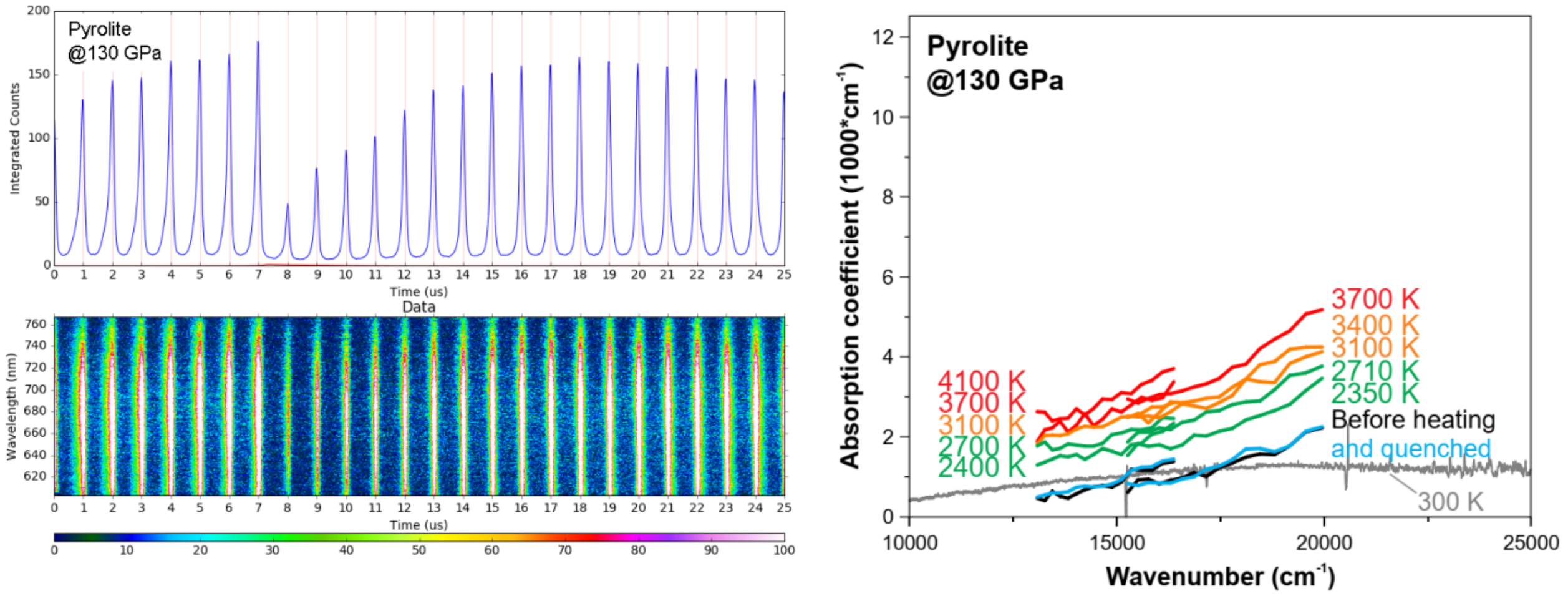

**Extended Data Figure 5. Left:** A streak camera image (bottom panel) and the corresponding integrated intensity (top panel) of pyrolite at 130 GPa. **Right:** Temperature-dependence of pyrolite absorption coefficients at 130 GPa (after applying scattering correction based on the 300 K absorption coefficients of Bgm6 at 117 GPa and Fp13 at 135 GPa).

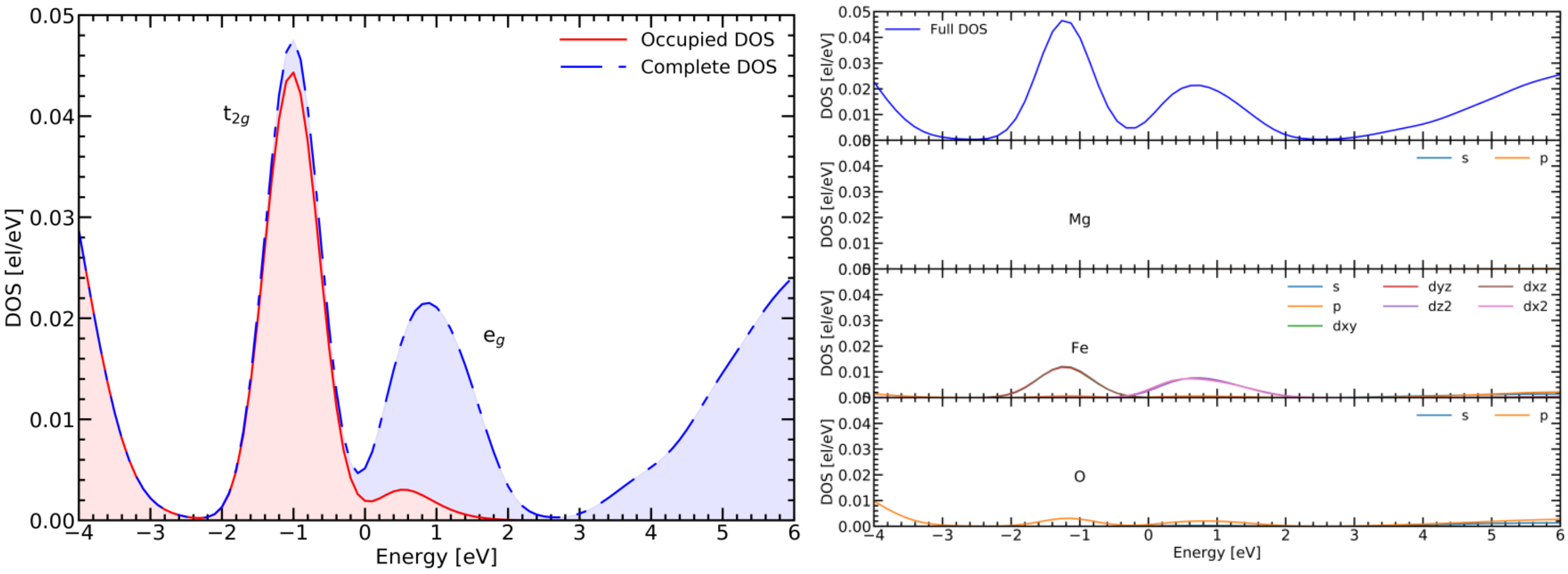

**Extended Data Figure 6.** Electronic density of states of 12.5 mol.% Fp at 135 GPa and 4300 K. The Fermi level is at 0 eV. **Left:** The blue curve is the complete DOS and the red one is the occupied DOS of Fe. **Right:** Element-projected DOS.

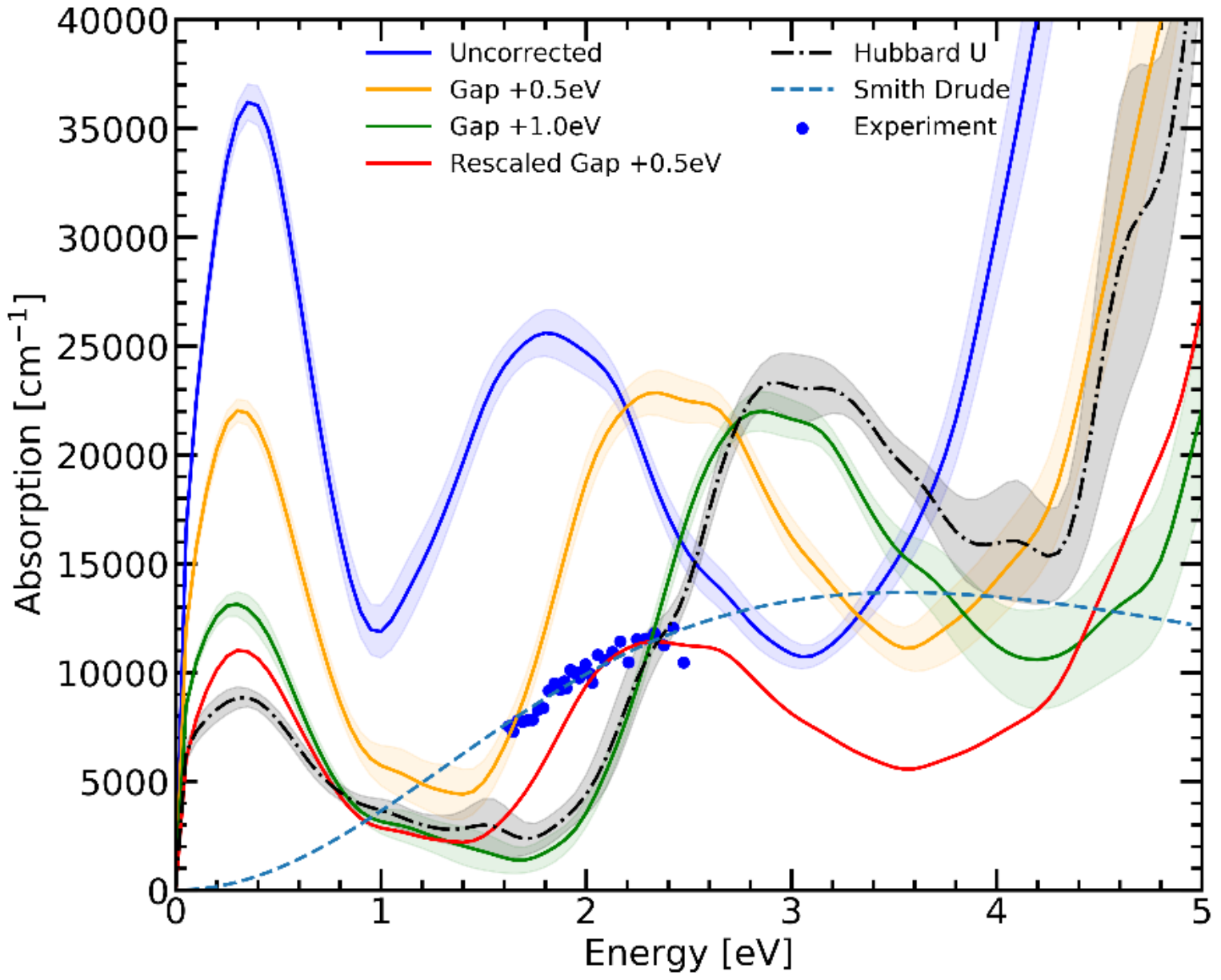

**Extended Data Figure 7.** Absorption coefficient of 12.5 mol.% Fp at 135 GPa and 4300 K as a function of the excitation energy. The blue curve is the direct result from Abinit. The orange and green curves are the results for the opened energy gap (up to 1 eV). The red curve is the +0.5eV gap rescaled by a factor of 0.5. The blue dots are the experimental results reported in Fig. 1B. The dashed line is a Smith-Drude fit to the experimental results.

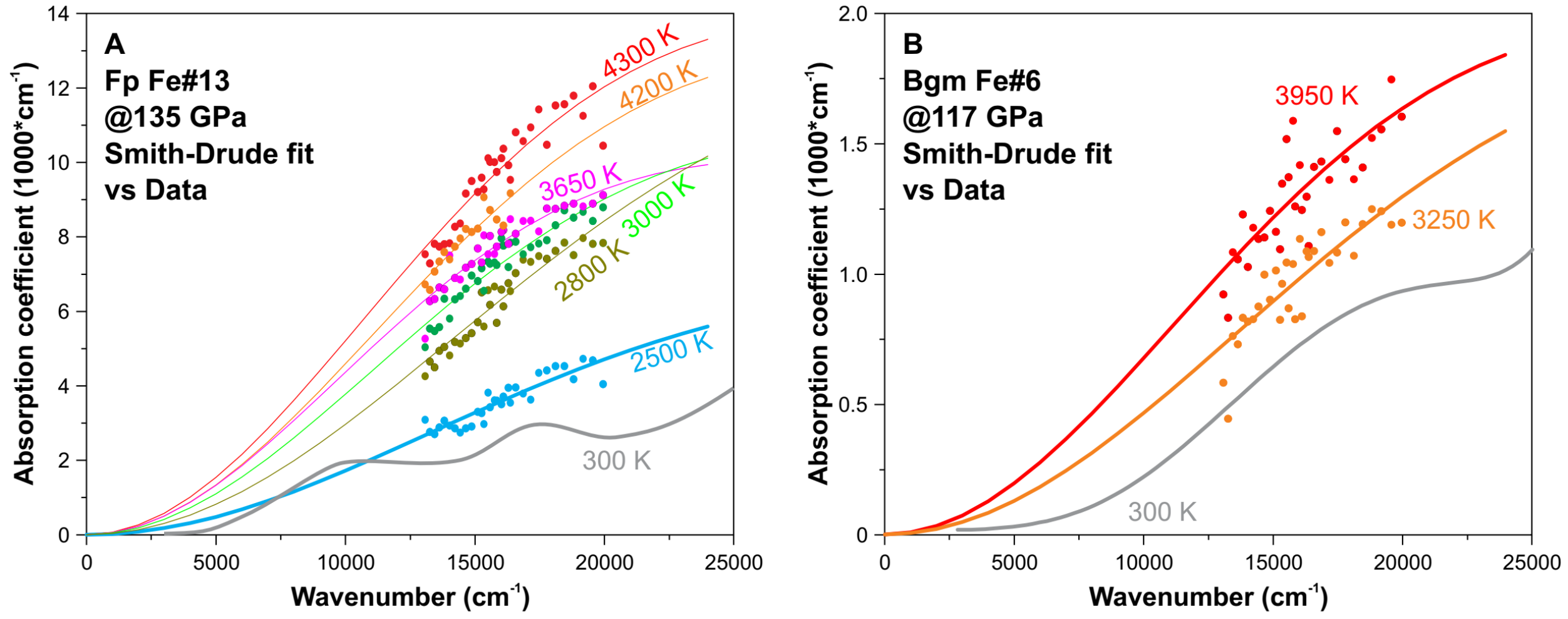

**Extended Data Figure 8.** Representative absorption coefficients of ferropericlase at 135 GPa (**A**) and bridgmanite at 117 GPa (**B**) as a function of temperature (circles, color-coded for temperature). Colored lines are Smith-Drude fits to the measured data. Grey lines are fits to wide spectral range absorption spectra measured at 300 K.

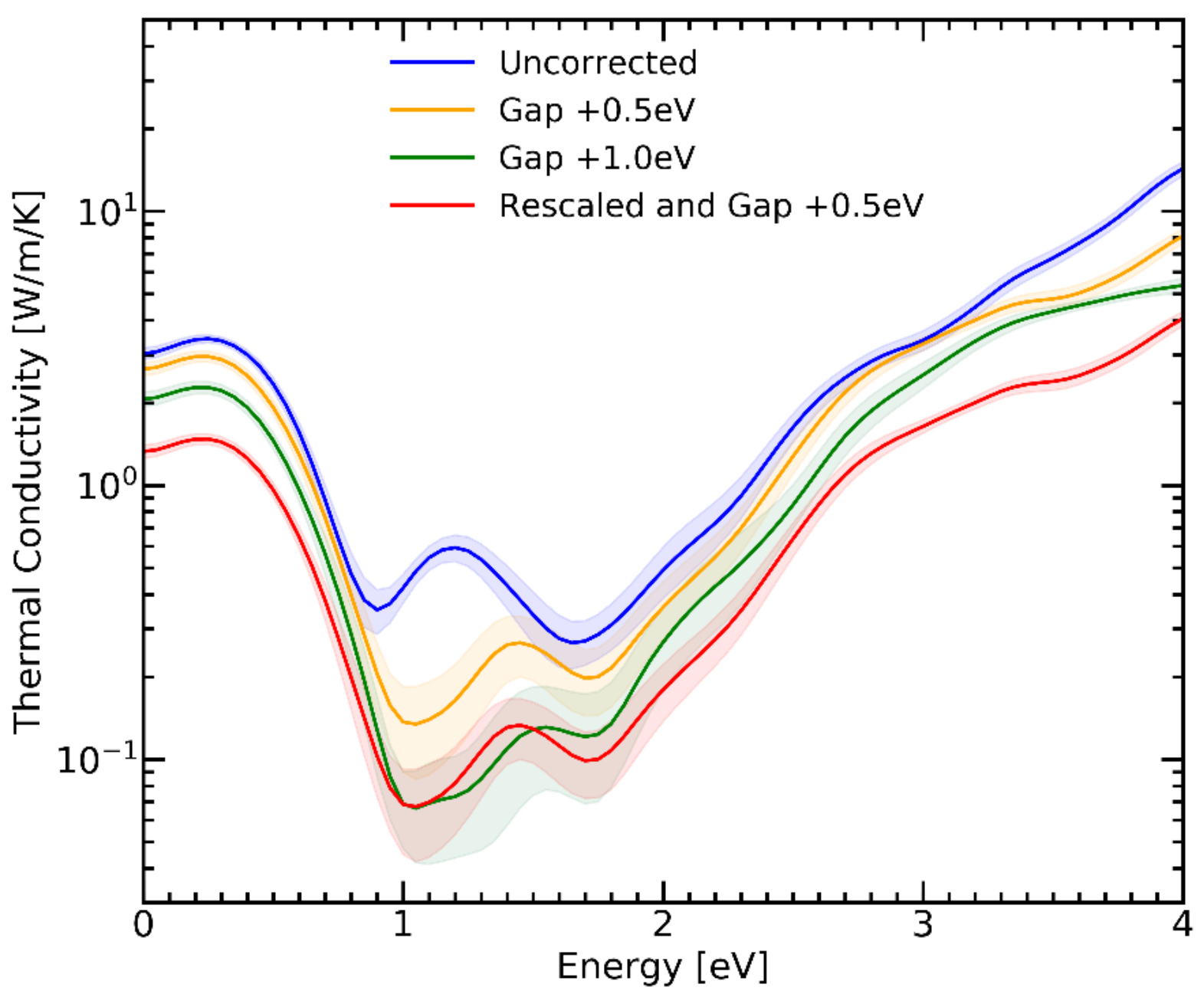

**Extended Data Figure 9.** Electronic contribution to the thermal conductivity of 12.5 mol.% Fp at 135 GPa and 4300 K as a function of frequency. The legend is similar to Extended Data Fig. 7.

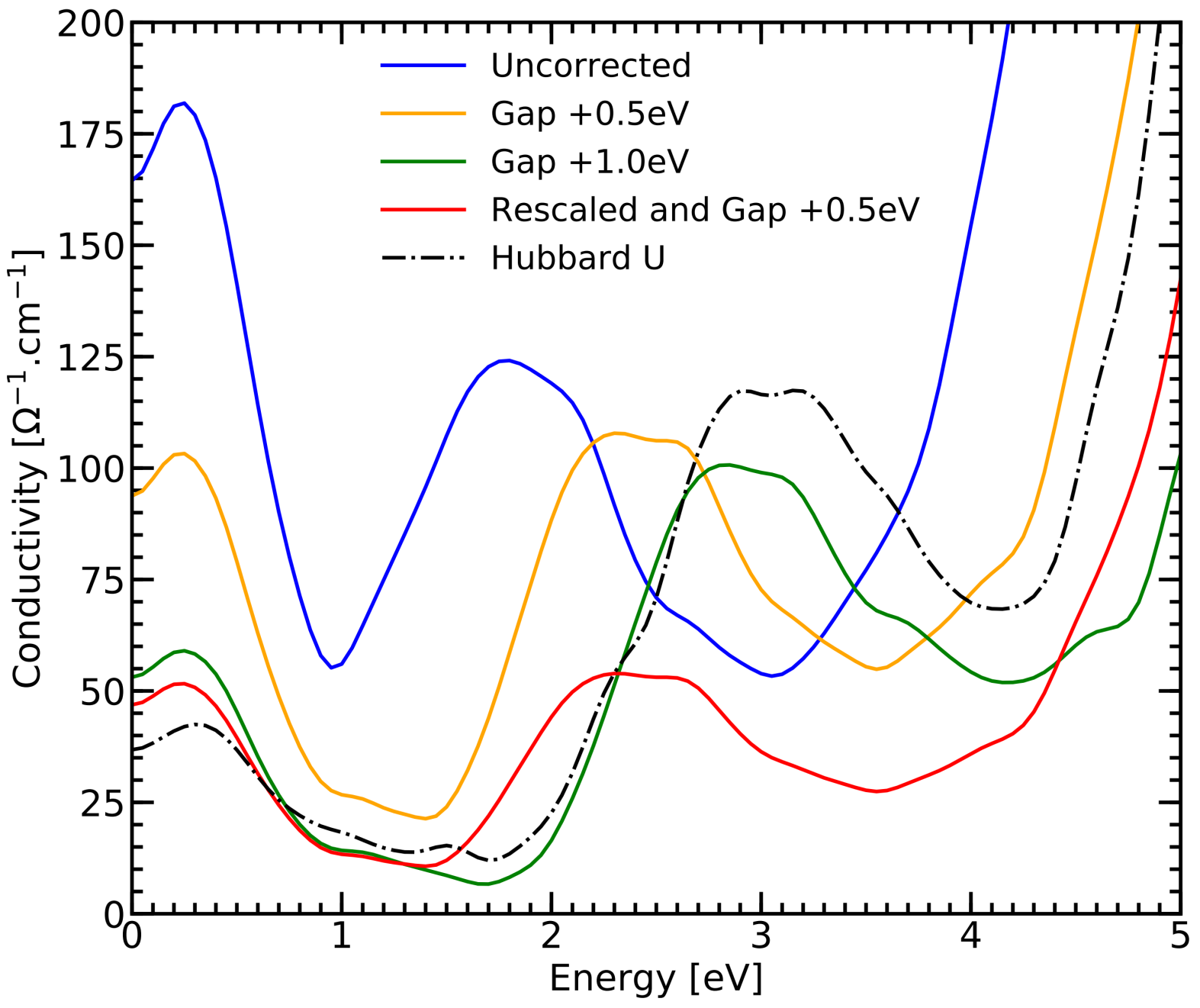

**Extended Data Figure10.** Electrical conductivity of 12.5 mol.% Fp at 135 GPa and 4300 K as a function of the excitation energy. The legend is similar to Extended Data Fig. 7.

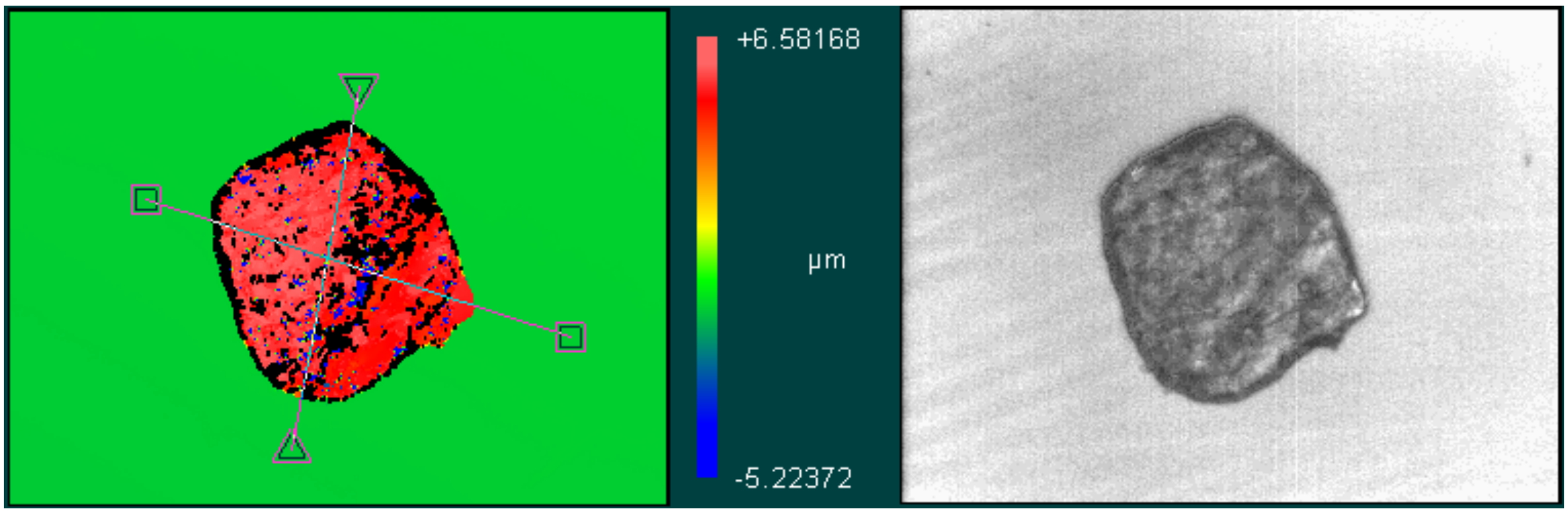

**Extended Data Figure 11.** ZYGO imaging of the bridgmanite sample used in this work after decompression from 117 GPa. The apparent thickness at 1 atm is 6.2 μm. **Left:** Thickness map. **Right:** Intensity map.

## Extended References

5 Goncharov, A. F., Haugen, B. D., Struzhkin, V. V., Beck, P. & Jacobsen, S. D. Radiative conductivity in the Earth's lower mantle. *Nature* **456**, 231-234 (2008).

6 Keppler, H., Dubrovinsky, L. S., Narygina, O. & Kantor, I. Optical absorption and radiative thermal conductivity of silicate perovskite to 125 Gigapascals. *Science* **322**, 1529-1532 (2008).

10 Goncharov, A. F., Beck, P., Struzhkin, V. V., Haugen, B. D. & Jacobsen, S. D. Thermal conductivity of lower-mantle minerals. *Phys. Earth Planet. Inter.* **174**, 24-32 (2009).

12 Goncharov, A. F. *et al.* Experimental study of thermal conductivity at high pressures: Implications for the deep Earth's interior. *Phys. Earth Planet. Inter.* **247**, 11-16 (2015).

17 Lobanov, S. S., Holtgrewe, N. & Goncharov, A. F. Reduced radiative conductivity of low spin $FeO_6$-octahedra in $FeCO_3$ at high pressure and temperature. *Earth Planet. Sci. Lett.* **449**, 20-25 (2016).

19 Holmstrom, E., Stixrude, L., Scipioni, R. & Foster, A. S. Electronic conductivity of solid and liquid (Mg, Fe)O computed from first principles. *Earth Planet. Sci. Lett.* **490**, 11-19 (2018).

21 Clark, S. P. Radiative transfer in the Earth's mantle. *Eos (formerly Trans. Am. Geophys. Union)* **38**, 931-938 (1957).

31 Akahama, Y. & Kawamura, H. Pressure calibration of diamond anvil Raman gauge to 310 GPa. *J. Appl. Phys.* **100**, 043516 (2006).

32 Syassen, K. Ruby under pressure. *High Pressure Res.* **28**, 75-126 (2008).

33 Goncharov, A. F. *et al.* Thermal equation of state of cubic boron nitride: Implications for a high-temperature pressure scale. *Phys. Rev. B* **75**, 224114 (2007).

34 McWilliams, R. S., Dalton, D. A., Mahmood, M. F. & Goncharov, A. F. Optical Properties of Fluid Hydrogen at the Transition to a Conducting State. *Phys. Rev. Lett.* **116**, 255501 (2016).

35 Lobanov, S. S., Goncharov, A. F. & Litasov, K. D. Optical properties of siderite ($FeCO_3$) across the spin transition: Crossover to iron-rich carbonates in the lower mantle. *Am. Mineral.* **100**, 1059-1064 (2015).

36 Lobanov, S. S., Hsu, H., Lin, J. F., Yoshino, T. & Goncharov, A. F. Optical signatures of low spin $Fe^{3+}$ in NAL at high pressure. *J. Geophys. Res.* **122**, 3565-3573 (2017).

37 Jiang, S. Q. *et al.* Metallization and molecular dissociation of dense fluid nitrogen. *Nat. Commun.* **9**, 2624 (2018).

38 Solomatov, V. S., El-Khozondar, R. & Tikare, V. Grain size in the lower mantle: constraints from numerical modeling of grain growth in two-phase systems. *Phys. Earth Planet. Inter.* **129**, 265-282 (2002).

39 Smith, N. V. Classical generalization of the Drude formula for the optical conductivity. *Phys. Rev. B* **64**, 155106 (2001).

40 Keppler, H., Mccammon, C. A. & Rubie, D. C. Crystal-field and charge-transfer spectra of (Mg,Fe)$SiO_3$ perovskite. *Am. Mineral.* **79**, 1215-1218 (1994).

41 Mermin, N. D. Thermal Properties of the Inhomogeneous Electron Gas. *Phys. Rev.* **137**, A1441-A1443 (1965).

42 Kohn, W. & Sham, L. J. Self-Consistent Equations Including Exchange and Correlation Effects. *Phys. Rev.* **140**, A1133-A1138 (1965).

43 Kresse, G. & Furthmuller, J. Efficient iterative schemes for ab initio total-energy calculations using a plane-wave basis set. *Phys. Rev. B* **54**, 11169-11186 (1996).

44 Nosé, S. A unified formulation of the constant temperature molecular dynamics methods. *J. Chem. Phys.* **81**, 511-519 (1984).

45 Nosé, S. Constant Temperature Molecular-Dynamics Methods. *Prog. Theor. Phys. Suppl.* **103**, 1-46 (1991).

46 Blochl, P. E. Projector augmented-wave method. *Phys. Rev. B* **50**, 17953-17979 (1994).

47 Gonze, X. *et al.* ABINIT: First-principles approach to material and nanosystem properties. *Comput Phys Commun* **180**, 2582-2615 (2009).

48 Monkhorst, H. J. & Pack, J. D. Special Points for Brillouin-Zone Integrations. *Phys. Rev. B.* **13**, 5188-5192 (1976).

49 Mazevet, S., Torrent, M., Recoules, V. & Jollet, F. Calculations of the transport properties within the PAW formalism. *High Energ. Dens. Phys.* **6**, 84-88 (2010).

50 Soubiran, F. & Militzer, B. Electrical conductivity and magnetic dynamos in magma oceans of Super-Earths. *Nat. Commun.* **9**, 3883 (2018).

51 Heyd, J., Scuseria, G. E. & Ernzerhof, M. Hybrid functionals based on a screened Coulomb potential. *J. Chem. Phys.* **118**, 8207-8215 (2003).

52 Kowalski, P. M., Mazevet, S., Saumon, D. & Challacombe, M. Equation of state and optical properties of warm dense helium. *Phys. Rev. B* **76**, 075112 (2007).

53 Ohta, K. *et al.* Experimental and Theoretical Evidence for Pressure-Induced Metallization in FeO with Rocksalt-Type Structure. *Phys. Rev. Lett.* **108** (2012).